\documentclass[prl,nofootinbib,superscriptaddress,twocolumn]{revtex4}
\usepackage[a4paper,left=1.5cm,right=1.5cm,top=3cm,bottom=3cm]{geometry}

\usepackage{amssymb,amsmath,amsfonts}
\usepackage[dvipsnames]{xcolor}
\usepackage{graphicx}
\usepackage{longtable}
\usepackage{verbatim}
\usepackage{color}
\usepackage{mdframed}
\usepackage{soul}
\usepackage{amsfonts,amssymb,mathrsfs,amsmath,esint}
\usepackage{slashed, cancel}
\usepackage{framed}
\usepackage{mdframed}
\usepackage{simplewick} 
\allowdisplaybreaks

\usepackage{latexsym}
\usepackage{graphicx}
\graphicspath{{./Figures/}}
\usepackage[dvipsnames]{xcolor}
\usepackage{booktabs}
\usepackage{datetime}
\newdateformat{mydate}{\THEDAY{ }\monthname[\THEMONTH]{ }\THEYEAR}

\usepackage{tikz}
\usepackage{color}
\usepackage{framed}
\usepackage{hyperref}
\hypersetup{colorlinks, citecolor=bluscuro, linkcolor=black, urlcolor=bluscuro}
\definecolor{rossos}{cmyk}{0,1,1,0.55}
\definecolor{bluscuro}{rgb}{0.15, 0.2, .85}
\definecolor{bluchiaro}{cmyk}{1,.3,0.,0.1}

\graphicspath{{./Figures/}}

\newcommand{\be}{\begin{equation}}
\newcommand{\ee}{\end{equation}}
\newcommand{\bea}{\begin{eqnarray}}
\newcommand{\eea}{\end{eqnarray}}
\newcommand{\beq}{\begin{equation}}
\newcommand{\eeq}{\end{equation}}
\def\beqa{\begin{eqnarray}}

\def\CMB{{\text{\tiny CMB}}}

\def\NL{\text{\tiny NL}}

\def\GW{{\text{\tiny GW}}}

\def\eeqa{\end{eqnarray}}

\def\lsim{\mathrel{\rlap{\lower4pt\hbox{\hskip0.5pt$\sim$}}
    \raise1pt\hbox{$<$}}}         
\def\gsim{\mathrel{\rlap{\lower4pt\hbox{\hskip0.5pt$\sim$}}
    \raise1pt\hbox{$>$}}}         

\usepackage[normalem]{ulem}

\begin{document}

\title{Anisotropies and non-Gaussianity of the Cosmological Gravitational Wave Background}

\author{N. Bartolo}
\address{Dipartimento di Fisica e Astronomia ``G. Galilei",
Universit\`a degli Studi di Padova, via Marzolo 8, I-35131 Padova, Italy}

\address{INFN, Sezione di Padova,
via Marzolo 8, I-35131 Padova, Italy}

\address{INAF - Osservatorio Astronomico di Padova, Vicolo dell'Osservatorio 5, I-35122 Padova, Italy}

\author{D. Bertacca}
\address{Dipartimento di Fisica e Astronomia ``G. Galilei",
Universit\`a degli Studi di Padova, via Marzolo 8, I-35131 Padova, Italy}

\address{INFN, Sezione di Padova,
via Marzolo 8, I-35131 Padova, Italy}

\author{S. Matarrese}
\address{Dipartimento di Fisica e Astronomia ``G. Galilei",
Universit\`a degli Studi di Padova, via Marzolo 8, I-35131 Padova, Italy}

\address{INFN, Sezione di Padova,
via Marzolo 8, I-35131 Padova, Italy}

\address{INAF - Osservatorio Astronomico di Padova, Vicolo dell'Osservatorio 5, I-35122 Padova, Italy}

\address{Gran Sasso Science Institute, Viale F. Crispi 7, I-67100 L'Aquila, Italy}

\author{M. Peloso}
\address{Dipartimento di Fisica e Astronomia ``G. Galilei",
Universit\`a degli Studi di Padova, via Marzolo 8, I-35131 Padova, Italy}

\address{INFN, Sezione di Padova,
via Marzolo 8, I-35131 Padova, Italy}

\author{A. Ricciardone}

\address{INFN, Sezione di Padova,
via Marzolo 8, I-35131 Padova, Italy}

\author{A. Riotto}

\address{Department of Theoretical Physics and Center for Astroparticle Physics (CAP)\\
			24 quai E. Ansermet, CH-1211 Geneva 4, Switzerland}

\address{CERN,
Theoretical Physics Department, Geneva, Switzerland}

\author{G.~Tasinato}
\address{Department of Physics, Swansea University, Swansea, SA2 8PP, UK}

\date{\today}

\begin{abstract}
\noindent
The Stochastic Gravitational Wave Background (SGWB) is expected to be a key observable for Gravitational Wave (GW) interferometry. Its detection will open a new window on early universe cosmology and on the astrophysics of compact objects.
Using a  Boltzmann approach, we study the angular anisotropies of the GW energy density, which is an important tool to disentangle the different cosmological and astrophysical contributions to the SGWB. Anisotropies in the cosmological background are imprinted both at its production, and by  GW  propagation through the large-scale scalar and tensor perturbations of the universe. The first contribution is not present in the Cosmic Microwave Background  (CMB) radiation (as the universe is not transparent to photons before recombination), causing an order one dependence of the anisotropies on frequency.  Moreover, we provide a new method to characterize the cosmological SGWB through its possible deviation from a Gaussian statistics. In particular, the SGWB will become a new probe of the primordial non-Gaussianity of the large-scale cosmological perturbations. 
\end{abstract}

\maketitle

\paragraph{Introduction.}

Operating ground based interferometers are not so far from reaching the sensitivity to detect the SGWB from unresolved astrophysical sources \cite{LIGOScientific:2019vic, Renzini:2019vmt}. On the other hand, future space-based, like LISA \cite{Audley:2017drz} and DECIGO \cite{Kawamura:2006up}, and earth-based, like Einstein Telescope \cite{Sathyaprakash:2011bh} and Cosmic Explorer \cite{Evans:2016mbw}, may be able to detect the stochastic background of cosmological origin, generated by early universe mechanisms of production of GWs \cite{Guzzetti:2016mkm,Bartolo:2016ami,Caprini:2018mtu,Geller:2018mwu,Ricciardone:2017kre,Dimastrogiovanni:2019bfl}. The most immediate way to differentiate the two backgrounds is from their frequency profile \cite{Caprini:2019pxz}. However, given that the SGWB is the sum of different contributions whose profiles are not fully known, it is important to develop also other means to characterize them. In this work we study the statistics of the angular anisotropies in the energy density of the GWs, that are either produced primordially or that are imprinted in the GWs as they propagate in the perturbed universe \cite{Alba:2015cms,Contaldi:2016koz, Bertacca:2017vod, Cusin:2017fwz, Jenkins:2018lvb, Cusin:2018avf}. 

This approach has several analogies with the well established formalism developed for CMB anisotropies. Following  \cite{Contaldi:2016koz}, we study, as commonly done for the CMB, the GW phase-space distribution function $f$, which can be immediately related to their energy density. We solve the collisionless Boltzmann equation for this distribution, and compute the 2-point and 3-point correlators of the GW energy density anisotropies on our sky. We focus on one crucial difference with the CMB: while the CMB temperature anisotropies are generated only at the last scattering surface, or afterward, the universe is instead transparent to GWs at all energies below the Planck scale. Therefore, the SGWB provides a snapshot of the universe right after inflation, and its anisotropies retain precious information about the primordial universe and the mechanisms for the GW formation. In particular, the primordial signal may be characterized by a significant (i.e. order one) dependence of the anisotropies on frequency. On the contrary, this dependence is very small in the CMB case, since any initial condition is erased by collisions before recombination, and any frequency dependence of the anisotropies is  generated only at second order in  perturbations   \cite{Bartolo:2006fj,Bartolo:2006cu,Chluba:2012gq}. We show, through a representative  example of sourced GWs during axion inflation  \cite{Cook:2011hg,Garcia-Bellido:2016dkw}, that a primordial GW signal visible at interferometer scales can indeed lead to anisotropies with large frequency dependence. 

Secondly, we study another important tool to characterise the cosmological SGWB, namely its non-Gaussianity. Recent works, starting with \cite{Bartolo:2018qqn},  investigated whether the GW 3-point function $\langle h^3 \rangle$ can be tested at interferometers. The measurement of this signal requires the measurement of phase correlations of the GW wave functions. As shown in  \cite{Bartolo:2018evs,Bartolo:2018rku}, two effects make such measurement unfeasible:  {\it (i)} the GW  propagation in the perturbed universe destroys any $\langle h^3 \rangle$ correlation possibly  present in the primordial signal; {\it (ii)} modes of nearby frequencies get confused with one another due to the finite  duration of the experiment, also resulting in a large phase decorrelation. There is, however, another type of non-Gaussianity that can be observed, and it is  the one present in the spatial distribution of the GW energy density. This does not involve initial phase correlation of the GW-field itself: here we present the first steps for the computation and study of the 3-point function (the bispectrum) of the GW energy density. 

For brevity reasons, this Letter contains only results under the simplest conditions. In a companion paper \cite{long} we shall present the details of these computations,  extend them to include the GW  propagation to second order in perturbations, as well as develop a more extended analysis of the GW bispectrum.

\noindent
\vskip 0.3cm
\noindent

\paragraph{Boltzmann equation for GWs.}
\noindent
We consider a distribution $f =  f (\eta ,x^i ,q ,{\hat n}^i ) $ of GWs as a function of their position $x^\mu$ and momentum $p^\mu =d x^\mu/d \lambda$, where $\lambda$ is an affine parameter along the GW trajectory. This distribution obeys the Boltzmann equation 
$\mathcal{L}[f] = \mathcal{C}[f(\lambda)] + \mathcal{I}[f(\lambda)]$,
where the Liouville term is $\mathcal{L}\equiv d/ d\lambda$, while $\mathcal{C}$ and $ \mathcal{I}$ account respectively for the collision of  GWs along their patch, and for their emissivity from cosmological and astrophysical sources \cite{Contaldi:2016koz}. The collision among  GWs affects  the distribution  at higher orders  (in an expansion series in the gravitational strength 
$1/M_{\it Planck}$) with respect to the ones we are considering, and can be disregarded. The emissivity can be due to astrophysical processes (such as black-holes merging) in the relatively late universe, as well as cosmological processes, such as inflation or phase transitions. In this work we are only interested in the stochastic GW background of cosmological origin, so we treat the emissivity term as an initial condition on the GW distribution (see \cite{Bartolo:2018igk} and Refs. therein for a discussion on collisional effects involving gravitons).   This leads us to study the free Boltzmann equation, $d f/d \eta = 0$ in the perturbed universe. Specifically, we consider scalar ($\Phi$ and $\Psi$) and tensor ($h_{ij}$, taken to be transverse and traceless) perturbations in the so-called Poisson gauge, around a homogeneous and isotropic background, giving the line element 
\begin{equation}
ds^2=a^2(\eta)\left[
-e^{2\Phi} d\eta^2+(e^{-2\Psi}\delta_{ij}+ h_{ij}) 
dx^i dx^j\right]\, ,
\label{metric}
\end{equation}
where $a ( \eta )$ is the scale factor, and $\eta$ is conformal time. 

Dividing the free Boltzmann equation by $p^0$ leads to 
\begin{equation}
\label{Dfc}
\frac{\partial f}{\partial \eta}+
\frac{\partial f}{\partial x^i} \frac{d x^i}{d \eta}+
\frac{\partial f}{\partial q} \frac{d q}{d \eta}+
\frac{\partial f}{\partial n^i} \frac{d n^i} {d \eta} = 0 \,, 
\end{equation}
where ${\hat n} \equiv {\hat p}$ is the direction of motion of the GWs, while $q \equiv \vert \vec{p} \vert a$ is the comoving momentum, that we use (as opposed to the physical one, that was used in  \cite{Contaldi:2016koz}, following the standard computation done for the CMB photons propagation \cite{Dodelson:2003ft}) as it simplifies the equation (\ref{Boltzmann-1st}) below. The first two terms in (\ref{Dfc}) encode free streaming, that is the propagation of perturbations on all scales.  At higher order this term also includes gravitational time delay effects. The third term causes the red-shifting of gravitons, including the Sachs-Wolfe (SW), integrated Sachs-Wolfe (ISW) and Rees-Sciama (RS) effects. The fourth term vanishes to first order and describes the effect of gravitational lensing. We shall refer to these terms as the free-streaming, redshift and lensing terms, respectively in a similar way to CMB physics.  Keeping only the terms up to first order in the perturbations, eq. (\ref{Dfc}) gives 
\begin{equation}
\frac{\partial f}{\partial \eta} + n^i \, \frac{\partial f}{\partial x^i} + \left[ \frac{\partial \Psi}{\partial \eta} - n^i \, \frac{\partial \Phi}{\partial x^i} + \frac{1}{2} \, n^i \, n^j \, \frac{\partial h_{ij}}{\partial \eta} \right] q \, \frac{\partial f}{\partial q} = 0 \;. 
\label{Boltzmann-1st}
\end{equation}

In analogy to the split in (\ref{metric}) we also assume that the GWs distribution has a dominant, homogeneous and isotropic contribution, with distribution function ${\bar f}$, plus a subdominant contribution $\delta f$. The two functions are obtained by solving eq. (\ref{Boltzmann-1st}) at zeroth and first order in  perturbations. Doing so, one immediately finds that any function ${\bar f} ( q )$ of the comoving momentum solves (\ref{Boltzmann-1st}) at zeroth order. 
As a consequence, the associated number density $n \propto \int d^3 p \, {\bar f} ( q )$ is diluted as $a^{-3}$ as the universe expands. This is also the case for  CMB photons, whose distribution function ${\bar f}_\CMB = (e^{p/T}-1)^{-1}$ is only controlled by the ratio  $p/T\propto p \, a = q$, where $T$ is the temperature of the CMB bath. This is  a consequence of the free particle propagation in an expanding background, and it does not rely on the distribution being thermal. 

 The subdominant anisotropic component $\delta f$ can be present as an initial condition. However, even if it is initially absent, eq. 
 (\ref{Boltzmann-1st}) shows that this anisotropy is produced by the propagation of the isotropic component ${\bar f}$ in the perturbed background. Assuming that $\partial \bar f/\partial q \neq 0$ (otherwise also the solution of $\delta f$ becomes trivial) 
it is convenient to rescale the perturbed part of the distribution function as 
\begin{equation}
\delta f \equiv  - q \, \frac{\partial {\bar f}}{\partial q} \, \Gamma \left( \eta ,\, \vec{x} ,\, q ,\, {\hat n} \right) \;. 
\label{Gamma-def}
\end{equation} 
In this variable and in Fourier space eq. (\ref{Boltzmann-1st}) gives 
\begin{equation}
\Gamma'+ i \, k \, \mu \Gamma = S (\eta, \vec{k}, {\hat n})  \, ,
\label{Boltfirstgamma1}
\end{equation}
where from now on prime denotes a derivative with respect to conformal time, $\mu$ is the cosine of the angle between $\vec{k}$ and ${\hat n}$, while the source function is $ S  = \Psi' - i k \, \mu \, \Phi -  \frac{1}{2}n^i n^j  \, h_{ij}' $. As we now show, the quantity $\Gamma$ can be immediately related to the anisotropic component of the GWs energy density, $\rho_\GW \equiv \int d^3 p \, p \, f $. It is customary to parametrize the GW energy density measured at the time $\eta$ at the location $\vec{x}$ in terms of its fractional contribution $\Omega_\GW$ through 
\begin{equation}
\rho_\GW \left( \eta ,\, \vec{x} \right) \equiv \rho_{\rm crit} \int d  \ln q \, \Omega_\GW \left( \eta ,\, \vec{x} ,\, q \right) \;, 
\end{equation}
where $\rho_{\rm crit} = 3 H^2 M_p^2$ is the critical energy density of the universe, and $H$ is the Hubble rate. Nearly all studies assume $\Omega_\GW$ to be homogeneous. Since we are interested in its inhomogeneous and anisotropic component, we have allowed $\Omega_\GW$ to depend on space. We account for the anisotropic dependence by defining $\omega_\GW$ through $\Omega_\GW = \int d^2 {\hat n} \, \omega_\GW ( \eta ,\, \vec{x} ,\, q ,\, {\hat n} )/4\pi$, and by introducing the density contrast $\delta_\GW \equiv  \delta \omega_\GW ( \eta ,\, \vec{x} ,\, q ,\, {\hat n} )/\bar \omega_\GW ( \eta ,\, q ) $. Using eq. (\ref{Gamma-def}), one then finds 
\begin{equation}
\delta_\GW  = \left[ 4 -  \frac{\partial \ln \, {\bar \Omega}_\GW \left( \eta ,\, q \right)}{\partial \ln \, q}  \right] \, \Gamma \left( \eta ,\, \vec{x} ,\, q ,\, {\hat n} \right) \,, 
\label{delta-Gamma}
\end{equation} 
with ${\bar \Omega}_\GW$  the homogeneous, isotropic component of $\Omega_\GW$. 

In the CMB case, by inserting the definition (\ref{Gamma-def}) in the Planck distribution, and expanding to first order, one finds $\Gamma_\CMB = \delta T/T$. The main difference between the CMB and the GW case is that, before recombination, the collision term between photons and baryons suppresses  any existing temperature anisotropy, thus removing any memory of the initial state. The observed temperature anisotropies $\delta T/T$ arise since recombination, following an equation analogous to (\ref{Boltfirstgamma1}), with a source that, to first order, is independent from the energy of the CMB photons. While in the CMB this dependence arises only to second order in perturbations, a significantly greater dependence can be present in the GWs distribution, as an initial condition.  In the following, we first compute and discuss the cosmological correlators of the GW anisotropies, and we then show through a concrete example that they can indeed have a significant dependence on frequency.

\vskip 0.3cm
\paragraph{Correlators of GW anisotropies and non-Gaussianity.}
\noindent
As it is standard \cite{Dodelson:2003ft}, we express each of the sources appearing in eq.\ (\ref{Boltfirstgamma1}) as a mode function times an initial variable that is constant at large scales, assuming for simplicity adiabatic scalar perturbations, and whose statistical properties have been set well before the propagation stage that we are considering (for instance during inflation, or during some early phase transition). Therefore, the scalar modes are (disregarding anisotropic stresses as for example those due to the relic neutrinos) $\Psi = \Phi \equiv T_\Phi ( \eta , k ) \, {\hat \zeta} ( \vec{k} )$; we then decompose the tensor modes as $h_{ij} \equiv \sum_{\lambda=\pm2} e_{ij,\lambda} ( {\hat k} ) h ( \eta , k )  {\hat \xi}_\lambda ( k^i )$, where the sum is over  right and left-handed  (respectively $\lambda=\pm2$)   circular polarizations, and the polarization operators are constructed as in \cite{Bartolo:2018qqn}. We insert these expressions in the source function in  (\ref{Boltfirstgamma1}), and solve for $\Gamma$. We then follow the treatment done for CMB perturbations, and we expand the solution in spherical harmonics, $\Gamma ( {\hat n} ) = \sum_\ell \sum_{m=-\ell}^\ell \Gamma_{\ell m} \, Y_{\ell m} ( {\hat n} )$, where we recall that ${\hat n}$ is the direction of motion of the GWs, and so the direction at which the GWs arrive on our sky. The multipoles $\Gamma_{\ell m}$ are the sum of three contributions. The first contribution arises from the initial conditions, 
\begin{equation} 
\frac{\Gamma_{\ell m,I} \left( q \right)}{4 \pi  \left( - i \right)^{\ell}} = 
\int \frac{d^3 k}{\left( 2 \pi \right)^3} \, 
\Gamma \big( \eta_{\rm in} ,\, \vec{k} ,\, q  \big)  \times Y_{\ell m}^* ( {\hat k}) \, j_{\ell} \big[ k \left( \eta_0 - \eta_{\rm in} \right) \big] \,,
\label{Glm,I}
\end{equation} 
where $\eta_0$ denotes the present time, and  we  set our location to $\vec{x}_0 = 0$. We  also remark that this term in general depends on $q$. The second contribution is due to the scalar sources in  eq. (\ref{Boltfirstgamma1}) 
\begin{eqnarray} 
\frac{\Gamma_{\ell m,S}}{4\pi  \left( - i \right)^{\ell}} &=&  \,  
\int \frac{d^3 k}{\left( 2 \pi \right)^3} \, 
\zeta ( \vec{k}) 
Y_{\ell m}^*( {\hat k}) \, {\cal T}_\ell^{(0)} (  k ,\, \eta_0 ,\, \eta_{\rm in} )  \,,\nonumber\\
&&
\label{Glm,S}
\end{eqnarray} 
where the scalar transfer function ${\cal T}_\ell^{(0)}$ is the sum of a term analogous to the SW effect for CMB photons, $T_\Phi ( \eta_{\rm in} , k ) \, j_\ell [ {k ( \eta_0 - \eta_{\rm in} )} ]$, plus the analog  of the ISW term, $ \int_{\eta_{\rm in}}^{\eta_0} d \eta' \, [  {T_\Psi'  ( \eta , k ) +  T_\Phi'  ( \eta , k ) }] \, j_\ell  [ {k ( \eta - \eta_{\rm in} )}]$. Finally, the third contribution $\Gamma_{\ell m,T}$ is due to the tensor modes in eq. (\ref{Boltfirstgamma1}), and it is formally analog to eq. (\ref{Glm,S}), with the product ${\hat \zeta} Y_{\ell m}^*$ replaced by the combination $\sum_{\lambda= \pm2}  {\hat \xi}_\lambda ( \vec{k} ) \, _{- \lambda}Y_{\ell m}^* ( \Omega_k ) $, involving the spin-2 spherical harmonics, and with the scalar transfer function replaced by the tensor one ${\cal T}_\ell^{(\pm 2)} ( k ,\, \eta_0 ,\, \eta_{\rm in} ) $, given by 
\be
{\cal T}_\ell^{(\pm 2)}  =\frac{1}{4}  \sqrt{\frac{\left(\ell +2 \right)!}{\left( \ell - 2 \right)!}}  \int_{\eta_{\rm in}}^{\eta_0} \!\!\!\! d \eta \,  h' \left( \eta ,\, k \right)   \frac{j_\ell \left[ k \left( \eta_0 - \eta \right) \right]}{ k^2 \left( \eta_0 - \eta \right)^2}. 
\ee
We are interested in  statistical correlators of the anisotropies. Under the assumption of statistical homogeneity and isotropy, the 2-point and 3-point correlators of ${\hat \zeta}$ are expressed in terms of, respectively, the scalar power spectrum and bispectrum through $\langle \zeta (  \vec{k})  \zeta^* (  \vec{k}' ) \rangle' =  (2 \pi^2/k^3) \, P^{(0)} (k) $ and  $\langle \zeta^3(  \vec{k}_i)  \rangle' =   B^{(0)} ( k_i)$ 
(we use the standard notation of the prime to eliminate the momentum conservation Dirac delta and the $(2\pi)^3$ coefficient).
  Analogously, correlators $P^{(\lambda)}$ and $B^{(\lambda)}$ can also be defined for the two tensor polarizations.  Moreover, we impose correlators of the same structure for the initial conditions,  namely $\langle \Gamma ( \eta_{\rm in} ,\, \vec{k} ,\, q )  \Gamma^* ( \eta_{\rm in} ,\, \vec{k}' ,\, q  ) \rangle' = (2 \pi^2/k^3) \, P^{ ( I )} ( k ) $ and  for the bispectrum $B^{(I)}$. In this work, we assume that the different contributions are uncorrelated.  Under these assumptions, one obtains $\langle \Gamma_{\ell m}   \Gamma_{\ell' m'}^*  \rangle \equiv \delta_{\ell \ell'} \, \delta_{mm'} \, {\widetilde C}_\ell =  \delta_{\ell \ell'} \, \delta_{mm'} [ {\widetilde C}_{\ell,I} ( q ) + {\widetilde C}_{\ell,S} + {\widetilde C}_{\ell,T} ] $, where we denote the correlators with a tilde to distinguish them from the CMB case. The contribution from the initial condition reads, 
\begin{equation}
\frac{{\widetilde C}_{\ell,I} \left( q \right)}{4 \pi} = \int \frac{d k}{k} \,  P^{(I)} \left( q ,\, k \right)   j_\ell^2 \left[k \left( \eta_0 - \eta_{\rm in} \right) \right]  \;, 
\end{equation} 
where again we stress the possible frequency dependence.  The other two terms are 
\begin{equation}
\frac{{\widetilde C}_{\ell,S} + {\widetilde C}_{\ell,T}}{4 \pi} = \sum_{\alpha=0,\pm2}  \int \frac{dk}{k} P^{(\alpha)} \left( k \right) \,  {\cal T}_\ell^{\left( \alpha \right) \,2}   \left( k ,\, \eta_0 ,\, \eta_{\rm in} \right) \;. 
\end{equation} 
At large scales, this contribution is dominated by the term proportional to the initial value of $\Phi$ in ${\cal T}_\ell^{(0)}$ (the analog  of the SW contribution for the CMB). For modes that re-enter the horizon during matter domination (as it is the case for those that give the large-scale anisotropies that we are considering), $T_\Phi = 3/5$ at early times \cite{Dodelson:2003ft}. So, for scale invariant power spectra, 
\begin{equation}
{\widetilde C}_\ell \simeq {\widetilde C}_{\ell,I} \left( q \right) +  {\widetilde C}_{\ell,S} \simeq \frac{2 \pi}{\ell \left( \ell + 1 \right)} \left[ P^{(I)} \left( q \right) + \left( \frac{3}{5} \right)^2 \, P^{(0)} \right] \;. 
\label{Cell}
\end{equation} 
The second term can be  compared to the SW contribution to the CMB anisotropies. In that case, the final temperature anisotropy is $1/3$ times the scalar perturbation at the last scattering surface, while $\Phi$ at that moment  decreased by a factor $9/10$ in the transition from radiation to matter domination  \cite{Dodelson:2003ft}. Therefore, ${\widetilde C}_{\ell,S} = (10/3)^2 C_\ell^{\rm SW}$. 

The structure of the bispectrum is forced by statistical isotropy to be a product of an $\ell_i$-dependent term times Gaunt integrals \cite{Bartolo:2004if}, $\langle \prod_{i=1}^3 \Gamma_{\ell_i m_i} \rangle \equiv {\widetilde b}_{\ell_1 \ell_2 \ell_3} \,   {\cal G}_{\ell_1 \ell_2 \ell_3}^{m_1 m_2 m_3}$. The initial condition term leads to 
\begin{eqnarray} 
{\widetilde b}_{\ell_1 \ell_2 \ell_3,I} &=&  \int_0^\infty d r \, r^2 \,  \prod_{i=1}^3 \Bigg\{ \frac{2}{\pi} \int d k_i \, k_i^2 \,  j_{\ell_i} \left[ k_i \left( \eta_0 - \eta_{\rm in} \right) \right] \nonumber\\ 
&& \quad\quad   \times  j_{\ell_i} \left( k_i \, r \right) \Bigg\} \, B^{(I)} \left( q ,\, k_1 ,\, k_2 ,\, k_3 \right) \;. 
\end{eqnarray} 
The scalar term ${\tilde b}_{\ell_1 \ell_2 \ell_3,S}$ is analogous, with the first spherical Bessel function replaced by the transfer function ${\cal T}_{\ell_i}^{(0)}$, and with the scalar bispectrum $B^{(0)}$ as last term. In particular, for a primordial bispectrum of the local form \cite{Komatsu:2001rj}, $B^ {(0)}  ( k_1 ,\, k_2 ,\, k_3 )  = (6 f_\NL/5) [(2 \pi^2)^2/(k_1^3 k_2^3) P^{(0)} ( k_1 ) P^{(0)} ( k_2 ) + 2 \; {\rm perm.} ] $, applying to the CMB result  \cite{Gangui:1993tt,Komatsu:2001rj}, the same rescaling done after eq. (\ref{Cell}) gives the dominant SW contribution at large scales 
\begin{equation} 
{\widetilde b}_{\ell_1 \ell_2 \ell_3,S} \simeq 2 \, f_\NL \left[ {\widetilde C}_{\ell_1,S}  {\widetilde C}_{\ell_2,S} +  2 \; {\rm perm.} 
 \right] \;. 
\end{equation}  
Finally, the tensor term reads 
\begin{eqnarray}
\frac{{\widetilde b}_{\ell_1 \ell_2 \ell_3,T}}{4 \pi}  &=&  
\, \left[ \prod_{i=1}^3 \int \frac{k_i^2 \, d k_i}{\left( 2 \pi \right)^3} {\cal T}_{\ell,i}^T (  k_i  ) \right]  \sum_{\lambda = \pm 2} 
\sum_{m_i} {\widetilde {\cal G}}_{\ell_1 \ell_2 \ell_3}^{m_1m_2m_3} \nonumber\\ 
&& \!\!\!\!\!\!\!\!  \!\!\!\!\!\!\!\!  \!\!\!\!\!\!\!\!  \!\!\!\!\!\! 
\times \left\langle \prod_{i=1}^3  \frac{4 \pi \left( - i \right)^{\ell_i}}{2 \ell_i+1} \int d \Omega_{k_i} \,   _{-\lambda}Y_{\ell_i m_i}^* \left( \Omega_{k_i} \right)  \, \xi_\lambda ( \vec{k}_i ) 
\right\rangle , 
\end{eqnarray} 
with the Wigner 3-$j$ symbols being employed in  defining 
\begin{eqnarray}
{\widetilde {\cal G}}_{\ell_1 \ell_2 \ell_3}^{m_1m_2m_3} \equiv \left( \begin{array}{ccc} 
\ell_1 & \ell_2 & \ell_3 \\ 
0 & 0 & 0 
\end{array} \right)^{-2} \, {\cal G}_{\ell_1 \ell_2 \ell_3}^{m_1m_2m_3} \,. 
\end{eqnarray}
For the case of purely adiabatic fluctuations, the formalism developed here allows us to determine   consistency relations for the squeezed limit of the bispectrum $B_{\delta}(\vec k_1,\,\vec k_2,\,\vec k_3) = \langle \delta_{\GW} ( \vec{k}_1 )  \delta_{\GW} ( \vec{k}_2 )  \delta_{\GW} ( \vec{k}_3 ) \rangle'$.  Such a squeezed limit  is determined by non-linear effects coupling long and short modes, and can be computed using well-known techniques developed in the context of cosmic inflation \cite{Maldacena:2002vr} and CMB \cite{Bartolo:2011wb,Creminelli:2011sq,Lewis:2012tc}. Focussing on a matter dominated universe, neglecting second-order tensor fluctuations,  and considering for simplicity only isotropic contributions to the density contrast, we find
\bea
&&\hskip-0.5cm \frac{ { B}_\delta(\vec k_1,\,\vec k_2,\,\vec k_3\to0) } {8 \pi^4}=\,
\left(\frac35\,\frac{\partial \ln \bar f(q)}{ \partial\,\ln q}\right)^3 \frac{{ P^{(0)}}(k_1)\,{ P^{(0)}}(k_3)}{k_1^3\,k_3^3} \nonumber
 \\
 &&\hskip1cm
 \times 
\left(
\frac{\partial \ln { P}^{(0)}(k_1)}{\partial\,\ln\,k_1} \, +
\frac25 
\frac{\partial \ln  q}{\partial   \ln \bar f( q)} \frac{\partial^2  \bar f( q)}{ \partial\,(\ln  q)^2}
 \right) .
\label{eqBisSq} 
\eea
Hence the squeezed limit of the bispectrum, in this specific situation,  depends on the tilt of scalar fluctuations, and  also on   derivatives of the background distribution function $\bar f(q)$, which is modulated by  long modes. We plan to further develop this subject in \cite{long}.

\begin{figure}[h!]
\centering 
\includegraphics[width=0.42\textwidth]{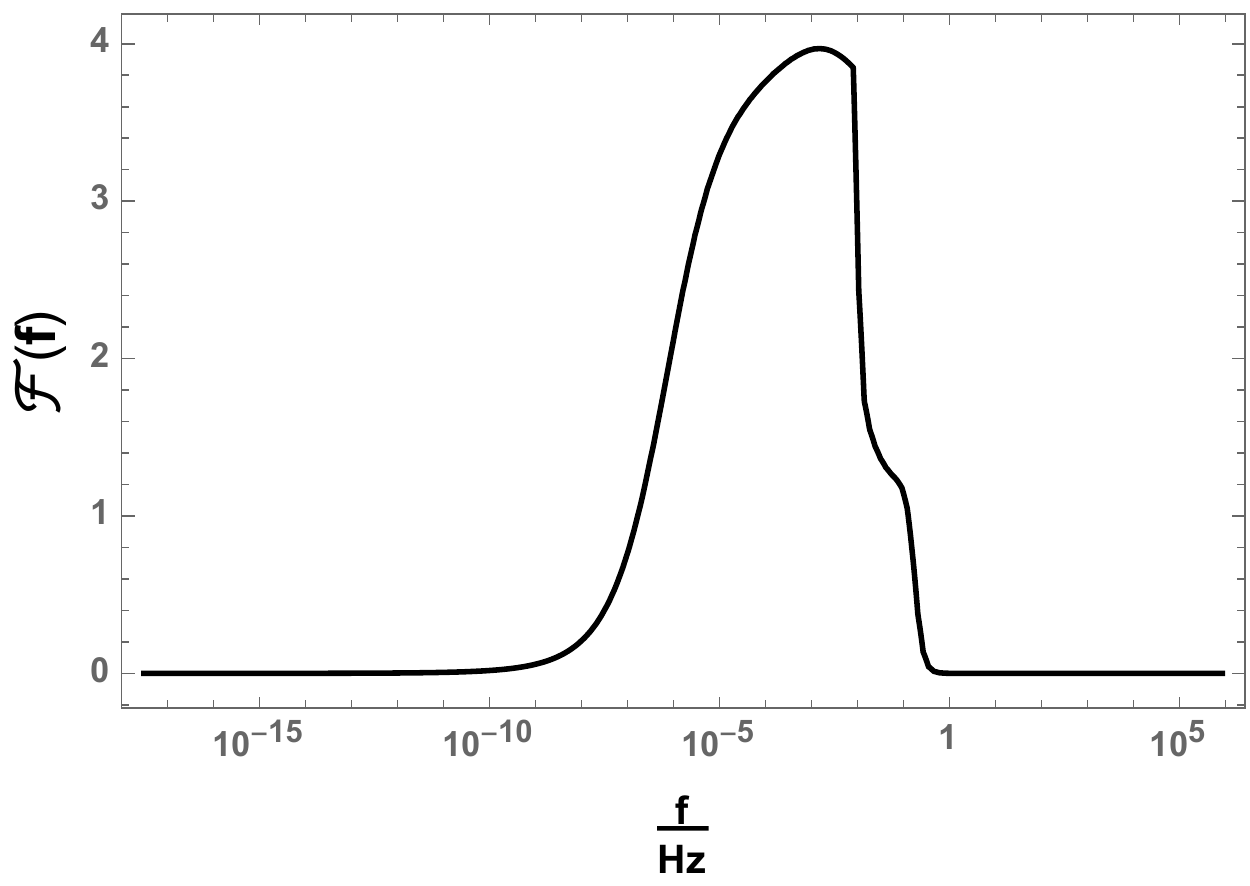}
\vskip -0.4cm
\caption{\it Quantity ${\cal F}$ as a function of the frequency $f = q/2 \pi$ of the GW signal for the model of axion inflation described in the text. 
}
\label{fig:F-sd}
\end{figure}

\vskip 0.1cm
\paragraph{Anisotropies in the Primordial SGWB and their frequency dependence.} 
Several mechanisms for the generation of a cosmological GW signal visible at interferometer scales have been studied in the literature \cite{Guzzetti:2016mkm,Bartolo:2016ami,Caprini:2018mtu}. Here we comment on a  specific mechanism where an axion inflaton $\phi$ sources gauge fields, which in turn generates  a large GW background. The amount of GWs sourced in this mechanism is controlled by the parameter $\xi \equiv (\dot{\phi}/2 f_\phi H)$, where $f_\phi$ is the decay constant of the axion inflaton. The inflaton background value then results in a background ${\bar \xi}$, and thus in a homogeneous and isotropic GW  component. The inflaton fluctuations result in a perturbation $\delta \xi$, and thus in the inhomogeneities of the primordial GW background. The anisotropy in the GW energy density arriving today at our location from a direction ${\hat n}$ is controlled by the value assumed by the parameter $\xi$ during inflation at the position $\vec{x}_0 + {\hat n} \, d$, where $d$ is the distance travelled by the GWs from their production during inflation to today.  GW modes observable at interferometers re-entered the horizon during the radiation-dominated era. The present fractional energy density $\Omega_\GW$ of these modes is 
equal to their primordial  power spectrum $P_\GW$ times a $q-$independent factor. Then, by  linearizing the primordial power spectrum in $\delta \xi$, the relation (\ref{delta-Gamma}) can be recast in the form $\Gamma_I  ( \eta_0 ,\, \vec{x}_0 ,\, q ,\, {\hat n} ) =  {\cal F} ( q ,\, {\bar \xi} ) \, \delta \xi  ( \vec{x}_0 + d \, {\hat n} )$, with 
\begin{eqnarray} 
{\cal F}  \equiv 
\left[  4 -   \frac{\partial \ln \left[  P_\GW \left( q ,\, {\bar \xi} \right) \right]}{\partial \ln \, q} \right]^{-1} \, \frac{\partial \ln \left[  P_\GW \left( q ,\, {\bar \xi} \right) \right]}{\partial {\bar \xi}} \,. 
\label{calF}
\end{eqnarray} 
We have then provided an immediate criterion for evaluating whether and how much the GW anisotropies depend on frequency  
(as, in principle, one could imagine a GW power spectrum for which the dependence on $q$ of  ${\cal F}$ vanishes, or is extremely suppressed). This conclusion only assumes that the primordial GW signal is function of some additional parameter $\xi$ which has small spacial inhomoegenities, and therefore it likely applies to several other mechanisms. For axion inflation, we consider the specific evolution shown in Figure 4 of \cite{Garcia-Bellido:2016dkw}, where the inflaton potential is chosen so to lead to a peak in the GW signal at LISA frequencies, without overproducing scalar perturbations and primordial black holes. We show in Figure \ref{fig:F-sd} the corresponding evolution of the parameter ${\cal F}$. We see that indeed this quantity presents a nontrivial scale dependence, and therefore the correlators of the anisotropies will be different at different frequencies. 
\paragraph{Future work.} We plan to extend the results presented here, to analyze several additional physical effects, including the effects of neutrinos on the GW amplitude \cite{Weinberg:2003ur}, the possible direct dependence of $\Gamma_I$ on ${\hat n}$, tests of non-standard expansion in the early universe, possible mixed bispectra among the three contributions to $\Gamma$ that we have discussed, and the feasibility of measuring the frequency dependence of the 2-point function and the bispectra at GW interferometers. 

\vskip 0.2cm
\noindent

\paragraph{Acknowledgments.}
\noindent
We thank Maresuke Shiraishi for useful comments and discussions. N.B., D.B. and S.M. acknowledge partial financial support by ASI Grant No. 2016-24-H.0.  A.R.~is supported by the Swiss National Science Foundation (SNSF), project {\sl The Non-Gaussian Universe and Cosmological Symmetries}, project number: 200020-178787.  The work of G.T. is partially supported by STFC grant ST/P00055X/1.


\end{document}